\documentclass[12pt]{iopart}
\pdfoutput=1 
\usepackage[T1]{fontenc}
\usepackage[utf8]{inputenc}
\usepackage{iopams, bbm}
\usepackage{math}
\usepackage[bottom,symbol]{footmisc}
\usepackage{scrextend}
\usepackage{enumerate, verbatim, hyperref, color, graphicx, titlesec}
	\graphicspath{{immagini/}}
\pdfpageattr{/Group << /S /Transparency /I true /CS /DeviceRGB>>}

\usepackage{harvard}
\usepackage{cite}
\begin{document}

\title[Allosteric and Cooperative models of E.\,coli motor]{%
%	A coarse-grained model for \emph{E.\,coli} flagellar motor
	From Conformational Spread to Allosteric and Cooperative models of \emph{E.\,coli} flagellar motor
}
\author{A. Pezzotta\,$^1$, M. Adorisio\,$^1$, A. Celani\,$^2$}

\address{$^1$ International School for Advanced Studies (SISSA), via Bonomea 265, I-34136 Trieste, Italy}
\address{$^2$ The Abdus Salam International Centre for Theoretical Physics (ICTP), Strada Costiera 11, I-34014 Trieste, Italy}
%\ead{mail...}
\vspace{10pt}
%\begin{indented}
%\item[]February 2014
%\end{indented}

\newcommand{\red}[1]{{\textcolor{red}{#1}}}
\newcommand{\eqref}[1]{(\ref{#1})}

\begin{abstract}
	Escherichia coli swims using flagella activated by rotary motors.
	The direction of rotation of the motors is indirectly regulated by the binding of a single messenger protein.
	The conformational spread model has been shown to accurately describe the equilibrium properties as well as the dynamics of the flagellar motor.
	In this paper we study this model from an analytic point of view. By exploiting the separation of time scales observed in experiments, we show how to reduce the conformational spread model to a coarse-grained, cooperative binding model.
	We show that this simplified model reproduces very well the dynamics of the motor switch.
	\end{abstract}

%\maketitle

%\section*{Outline}
%	
%
%Introduction
%\begin{enumerate}
%\item The bacterial motor
%\item Conformational spread model vs MWC (also Alon)
%\item Time-scale separation -> motor as a cooperative binding unit
%\end{enumerate}
%Conformational spread
%\begin{enumerate}
%\item variables, independent binding
%\item Glauber dynamics
%\item intractability in general
%\end{enumerate}
%From CS to MWC
%\begin{enumerate}
%\item Strong coupling $J$
%\item domain walls  and heuristics
%\end{enumerate}
%From MWC to cooperative binding
%\begin{enumerate}
%\item Fast  act/deact wrt to binding
%\item discuss effective cooperativity /birth-death
%\end{enumerate}
%Results for effective dynamics
%\begin{enumerate}
%\item Eq.
%\item Switching times (no barrier/diffusive like see Duke)
%\end{enumerate}
%Discussion: time-scale separation, other motors

	\section{Introduction}
	
	The ability to efficiently respond to chemical stimuli is essential for the survival of many animal species, ranging from prokaryotic cells to much more complex organisms such as insects or birds.
	At the microscale, the mechanism which allows organisms to move under the influence of chemical stimuli is called \emph{chemotaxis} \cite{berg}.
	
	Escherichia coli (\emph{E.\,coli}) is one of the model organisms for studies about bacterial chemotaxis \cite{berg_03}.
	Thanks to its flagella, activated by bi-directional rotary motors, \emph{E.\,coli} is able to move towards more favorable environments by optimally alternating \emph{runs} and \emph{tumbles}, which approximately consist of straight lines and random ``turns'', respectively.
	
	The biochemical mechanisms underlying the chemotactic response of \emph{E.\,coli} are well understood at the molecular level \cite{sb_02}.
	A sensing apparatus is devoted to detecting information about the environment, by measuring concentration of chemicals (generally called, in this context, \emph{chemoeffectors}).
	The arrangement and functioning of the receptors present on \emph{E.\,coli} cellular membrane has been extensively investigated also from the theoretical point of view (see, \eg \cite{shi_98,shimizu_00}).
	The information collected by the receptors is transduced to the flagellar motors through the ``messenger molecule'' CheY.
	The cytoplasmic concentration of its phosphorylated form CheY-P varies according to the activity of the membrane receptors.
	The CheY-P molecule then acts as a regulator of the activity of the flagella by binding to their motors.
	These are constituted by rings of Fli molecules, arranged in units called \emph{protomers}.
	Motors are biased by the Fli occupancies to rotate counterclockwise (CCW) or clockwise (CW).
	When all the motors are in the CCW state, flagella form a bundle which propels the cell in a forward run; if at least one motor is in the CW state instead, the bundle splits apart and the cell tumbles.

	Such mechanism is an example of \emph{allosteric} (or \emph{indirect}) regulation, where the activity of protein complexes changes collectively upon independent binding of external molecules.
%	Some models of this kind of regulation have been proposed in the last decades, all essentially inspired by the seminal works by Monod, Wyman and Changeux \cite{mwc_65} and by Koshland, Nemethy and Filmer \cite{knf_66}, the \emph{concerted} (MWC) and \emph{sequential} (KNF) model, respectively.
	The original model which encodes the concept of cooperativity in indirect regulation is the one proposed by Monod, Wyman and Changeux (MWC), commonly known as \emph{concerted} model \cite{mwc_65, mgp_13}.
	
	Shortly after the paper by Monod, Wyman and Changeux, Eigen realized that the concerted model can be extended in order to offer a more graded interplay between the interactions within allosteric complexes and their binding affinities \cite{eigen_67}.
	When the interactions are local, this generalized model takes the name of \emph{conformational spread} model (see Sec.\,\ref{sec:CS}) and is nowadays understood in a statistical mechanical framework in the light of the ferromagnetic Ising model, to which it is formally equivalent~\cite{shi_98, db_01}.
	
%	Later, Eigen \cite{eigen_67} proposed a more general picture in which the concerted and sequential models figure as extreme limits: this model has been later on given the name of \emph{conformational spread} model.
	
	These allosteric models have found application in bacterial chemotaxis.
	In Ref.\,\cite{alon_98}, the authors showed how the MWC model is able to reproduce the activity of the flagellar motor of \emph{E.\,coli} as a function of the concentration of cytoplasmic CheY-P. \cite{alon_98}.
	In this paper, the authors recognized that the balance between the different CheY-P affinity in the two activity states and the size of the motor protein complex was essential in explaining the observed cooperative behaviour of the switch.
	The MWC model turned out to be particularly suitable for describing the flagellar switch of \emph{E.\,coli}, in that it accounts for the correct degree of cooperativity with a proper choice of the parameters.

	The conformational spread model has been applied to bacterial chemotaxis, both for the membrane receptors \cite{shi_98, db_01} and for the flagellar rotary motors \cite{dnb_01}.
	By means of a simulation of its associated Glauber dynamics \cite{redner_kin}, a numerical test of the conformational spread model against the experimental measurement of the rotation speed of the flagella has been performed \cite{bai_10}.
	Such analysis showed an excellent agreement between experiments and numerical simulations regarding several aspects of the dynamics, such as the switching time distribution at fixed values of the cytoplasmic CheY-P concentration and the sensitivity of the switch upon small variation of CheY-P.
	A more detailed numerical analysis of the model followed up \cite{ma_12}, in which also other dynamical properties of the conformational spread model were quantified (like the locked-state behaviour, namely, the time spent by the motor in a rotational state between two consecutive switches) and a more precise estimation of the parameters of the model which best fit the experimental results was given.
	
	From the analytical point of view, one major obstacle to the study of the conformational spread model resides in the large number of states.
	The one-dimensional nature of the ring allows nonetheless for an exact calculation of its partition function at equilibrium via the transfer matrix method \cite{mmr_10}.
	However, no analytical treatment of the non-equilibrium behaviour of the model has ever been attempted, to our knowledge.
	
	In this work we present an analytical derivation of the non-equilibrium properties of the conformational spread as a model of the flagellar switch.
	Our analysis hinges upon the presence of a hierarchy of widely separated time scales, as confirmed by experiments.
	Due to the strong interaction between the protomers, the coarsening of activity domains in the ring is much faster than the nucleation of a domain, \emph{i.e.} the transitions away from the state of all active or all inactive protomers. This allows the treatment of the whole motor as an allosteric switch in two different activity states (CW and CCW), essentially described by the MWC model.
	The nucleation of a domain is in turn much more frequent than the binding/unbinding of a CheY-P molecule by one protomer, which makes it possible to operate a quasi-static approximation for the number of bound CheY-P and get a description of the slow binding dynamics, to which the activity is slaved.
	This separation of time scales allows us to reduce the complexity of the full conformational spread dynamics by progressively averaging the faster degrees of freedom and obtain, in the end, an effective cooperative model which captures the relevant features of the flagellar switch on the slowest time scales.
	The effective rates of the emergent ``coarse-grained'' cooperative binding model are expressed in terms of the rates of the original ``microscopic'' conformational spread model.
	In short, the rationale of our approach can be schematically summarized as follows:
	\begin{center}
		\includegraphics[width=\textwidth]{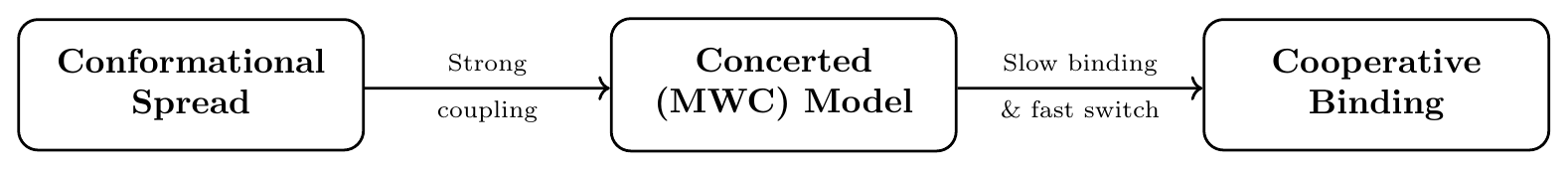}
	\end{center}
	
	The paper is structured as follows:
	in Sec.\,\ref{sec:CS} we present the conformational spread model, outlining its equilibrium properties and introducing the dynamics (satisfying detailed balance) which is relevant for our study and is the object of our multiscale analysis;
	in Sec.\,\ref{sec:2MWC} we show that, in our experimentally justified assumptions, it is possible to reduce the conformational spread to the concerted MWC model;
	a further time-scale separation is the subject matter of Sec.\,\ref{sec:2BD}, resulting in a cooperative binding model (formally, a birth-and-death process with site-dependent rates) that is compared with experiments in Sec.\,\ref{sec:BD}.

	\section{Conformational Spread Model}\label{sec:CS}
	
\begin{figure}[t]
		\centering
%		\small\def\svgwidth{.5\textwidth}
%		\input{immagini/CS_configs.pdf_tex}
		\includegraphics[scale=1]{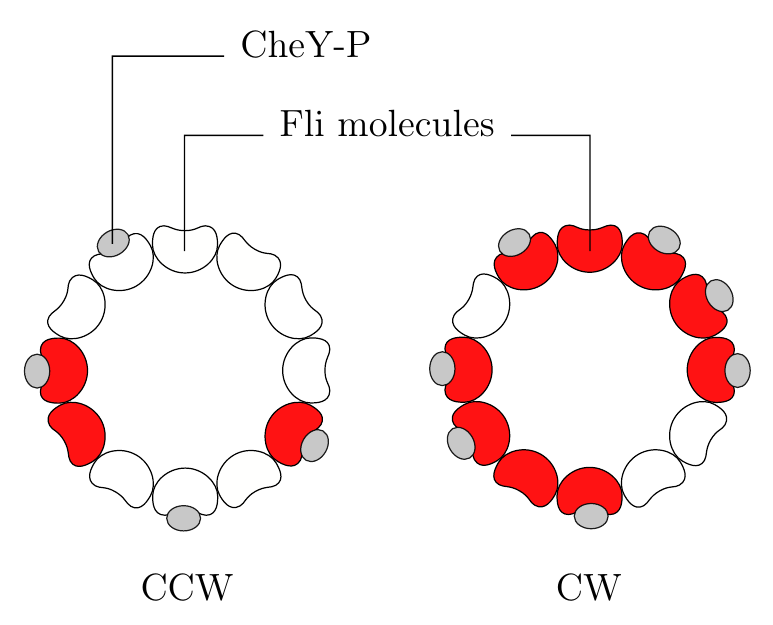}
		\caption{\textbf{The flagellar motor.}
			The Fli molecules are depicted in white (inactive state, $I$) and red (active state, $A$), while the grey spots represent the CheY-P regulator.
			The motor rotates counterclockwise when most of the protomers are in the inactive state (left) and clockwise otherwise (right).
		}\label{fig:motor-configs}
	\end{figure}

	The ring of proteins forming the motor of the \emph{E.\,coli} flagella has been shown to be very well described by the conformational spread model \cite{db_01, dnb_01, bai_10}.
	This model consists in $N$ identical units, or \emph{protomers}, each of which can appear in two different states, active ($A$) or inactive ($I$):
	a protomer in the active state increases the probability of CW rotation of the motor, and of CCW rotation in the inactive state (see Fig.\,\ref{fig:motor-configs}).
	Moreover, each protomer can also bind a \emph{ligand}, corresponding to the CheY-P chemotactic regulator: we refer to the protomer as in the \emph{bound} ($B$) state when a ligand is attached to it, or \emph{unbound} ($U$) otherwise.
	Therefore, the single protomers can be in 4 different states, corresponding to all the possible \emph{activity} and \emph{binding} configurations.
	
\begin{figure}[t]
		\centering
%		\small\def\svgwidth{.8\textwidth}
%		\input{immagini/CS_states_coupling.pdf_tex}
		\includegraphics[scale=1]{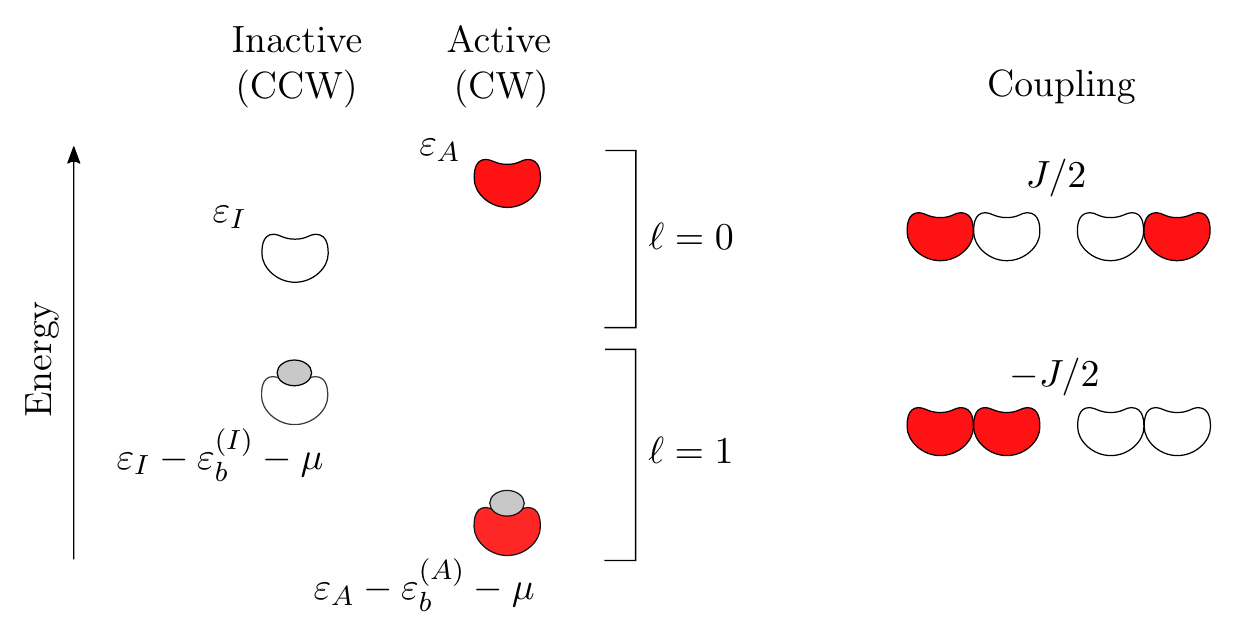}
		\caption{\textbf{State diagram and couplings in the Conformational Spread Model.}
			On the left, the energy levels of the single protomer states: the active (CW) configuration is energetically favorable in the unbound case ($\ell= 0$), while the inactive (CCW) has lower energy when in the bound case ($\ell=1$); the binding regulates the activity of the protomers. The notation and the general scheme has been borrowed from \cite{mgp_13}.
			On the right, the coupling energy: the ``ferromagnetic'' coupling (independent of $\ell$) accounts for the high sensitivity of the response of the ring upon binding.
		}\label{fig:CS-states}
	\end{figure}

	The state diagram of a single protomer is depicted in Fig.\,\ref{fig:CS-states}:
	the $A$ state is energetically more favorable than the $I$ state when a ligand is bound and vice versa.
	This property ensures that this is a good model for allosteric regulation. Namely, a bias in the activity of the motor depends on the number of bound CheY-P molecules:
	at fixed high concentration of cytoplasmic CheY-P (denoted by $c$) the motor will most probably spin clockwise.
	The state of the full system is specified by the sequence $s = \{(\alpha_1,\,\ell_1),\,\ldots (\alpha_N,\,\ell_N)\}$, where the subscripts label the $N$ protomers, $\alpha$ indicates the activity state $A$ or $I$, and $\ell$ stands for the binding state $B$ ($\ell = 1$) or $U$ ($\ell=0$): hence, the number of possible configurations of the ring with $N$ protomers is $(2\times 2)^N$.
	
	In addition, the protomers are coupled via a nearest neighbour interaction, which depends on their activity states \emph{only}: in particular, the energy is lowered by a quantity $J$ when the neighbouring protomers are in the same activity state $A$ or $I$.
	It turns out that the activity of the ring (fraction of active protomers) is more sensitive to small variations of concentration of ligands in the interacting case than in a system of $N$ independent protomers.
	Therefore, the coupling is an essential ingredient which enhances the sensitivity of the whole complex.
	
	The conformational spread model is very reminiscent of the Ising model. %, or, more generally, the Potts model.
	In fact, if one associates to each protomer a \emph{spin} variable $\sigma_i$ taking value $+ 1$ when the protomer is active ($\alpha_i = A$), or $-1$ when it is inactive ($\alpha_i= I$), one can represent the states of the system as $s = \{(\sigma_i, \ell_i)\}_{i=1}^N$ and the equilibrium properties of the model are determined by the Hamiltonian
	\begin{equation}\label{eq:CS-ham}
		H = - \frac{J}{2} \sum_{\av{i,\,j}} \sigma_i\,\sigma_j - \sum_i h(\sigma_i,\,\ell_i) \comma
	\end{equation}
	where $J$ is a positive constant and $h$ is the single-protomer contribution, reproducing the energy diagram in Fig.\,\ref{fig:CS-states},
%	\begin{equation}
%		\begin{aligned}
	\begin{eqnarray}\label{eq:CS-energies}
			h(\sigma,\,\ell) =
%			- \varepsilon_\alpha + (\varepsilon_b^{(\alpha)} + \mu)\,\ell \comma
			\frac{1}{2}\bigg[\varepsilon_I - \varepsilon_A - &(\varepsilon_b^{(I)} - \varepsilon_b^{(A)})\ell \bigg]\sigma \nonumber\\
				&- \frac{1}{2}\bigg[\varepsilon_I + \varepsilon_A - (\varepsilon_b^A + \varepsilon_b^I + 2\mu )\ell  \bigg] \fs
%		\end{aligned}
%	\end{equation}
	\end{eqnarray}
	The one in Eq.\,\eqref{eq:CS-ham} is an Ising Hamiltonian with ferromagnetic coupling $J$, where $h$ plays the role of an external local magnetic field, set by the occupation $\ell$;
	in Eq.\,\eqref{eq:CS-energies}, $\mu$ is the chemical potential, determined by the concentration of CheY-P, $c$, by
	\begin{equation}\label{eq:chempot}
		\mu = \mu_0 + \frac{1}{\beta}\,\ln \frac{c}{c_0} \comma
	\end{equation}
	where $\mu_0$ and $c_0$ are reference chemical potential and concentration, respectively.
	Hereafter, the notation $\sigma$ and $\alpha$ will be used interchangeably, according to the situation.
	The partition function $Z = \sum \exp(-\beta H(s))$  (where $\beta = 1/k_B T$ is the inverse temperature and the sum is done over the $4^N$ possible states of the ring of protomers) has been calculated exactly via transfer matrix approach \cite{mmr_10}.
	The analytic results found therein fit very well the experimental curves \cite{sb_02} of the ligand occupancy (average fraction of bound protomers) and the activity (fraction of protomers in the $A$ state) as a function of the concentration of CheY-P.\\
	
	If on one hand the equilibrium properties of the conformational spread model are exactly known, on the other hand a full-fledged analytic treatment of the stochastic dynamics of this model seems difficult.
%	In principle, there is not a unique way of defining the dynamics which leads to the thermal equilibrium specified by Eq.\,(\ref{eq:CS-ham}) given an initial configuration: 
%
	In the definition of the conformational spread model given above, there is no prescription about the dynamics.
	A natural choice which satisfies detailed balance is the Glauber-like \cite{redner_kin} Markovian dynamics, used in numerical simulations of this model in Refs.\,\cite{bai_10, ma_12}.
	In such prescription, the process $\{S_t\}_t$ which accounts for the kinetics of the conformational spread model is governed by the master (Kolmogorov) equation
	\begin{equation}\label{eq:CS-glauber-master}
		\frac{\partial}{\partial t} P(s,t) = \sum_{s'} \left[ P(s',t)\,K(s'\to s) - P(s,t)\, K(s\to s')\right] \comma
	\end{equation}
	where $P(s,t) = \mbox{Prob}\{S_t=s\}$ and $K$ are the rates defined as
	\vspace{1ex}
	\begin{equation}\label{eq:CS-glauber}
		\eqalign{
		K(s\to s') = \bigg\{\frac{\omega_f}{1-\gamma}& \Big( 1 - \gamma\,\sigma_i\,\frac{\sigma_{i+1} + \sigma_{i-1}}{2} \Big)\,e^{\beta\,h(-\sigma_i,\,\ell_i)} \,\delta_{\sigma_i',\,-\sigma_i}\,\delta_{\ell_i',\,\ell_i} \\
				&+     \omega_s\,e^{\beta\,h(\sigma_i,\,1-\ell_i)}\,\delta_{\sigma_i',\,\sigma_i}\,\delta_{\ell_i',\,1-\ell_i}\bigg\}\,\prod_{j\neq i}\delta_{\sigma_j',\,\sigma_j}\,\delta_{\ell_j',\,\ell_j} \comma
		}
	\end{equation}
%	\vspace*{1.5ex}
	where the product of Kronecker $\delta$ indicates that the rates $K$ only involve one protomer at a time.
	Each term in Eq.\,\eqref{eq:CS-glauber} is obtained from detailed balance up to multiplicative factors $\omega_f$ and $\omega_s$:
	these constants account for typical time scales of the flipping and binding process, respectively.
	The constant $\gamma$ in the spin-flip contribution is set by the strength of the coupling, $\gamma = \tanh(\beta\,J)$.
	\begin{figure}[b]
		\centering
		\includegraphics[scale=1]{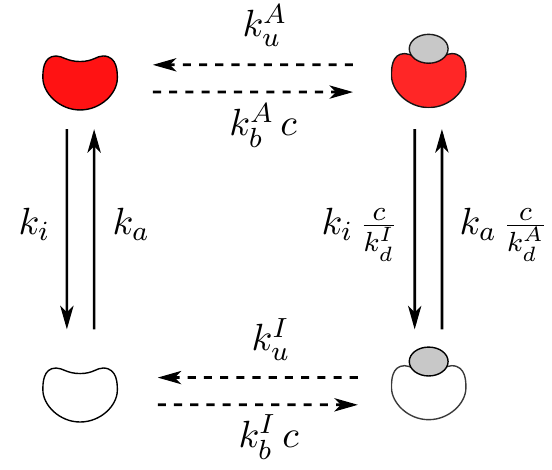}
		\caption{\textbf{Single-protomer dynamics.} Contributions to the transition rates $K^{(i)}$ from the single-body term in the Hamiltonian \eqref{eq:CS-ham}: vertical arrows are labeled by the rates of activation/inactivation; horizontal arrows by the binding/unbinding ones.}
		\label{fig:CS-single}
	\end{figure}
	For a system made of a one protomer (or for a single protomer in absence of interaction, $\gamma = 0$), according to Eq.\,(\ref{eq:CS-glauber}), we define the constants $k_a$ and $k_i$ as the rates for activation and inactivation with $\ell = 0$,
	\begin{equation}
		k_a = \omega_f\,e^{-\beta\varepsilon_A} \comma
		\qquad \mbox{ and } \qquad
		k_i = \omega_f\,e^{-\beta\varepsilon_I} \smc
	\end{equation}
	their counterparts for $\ell = 1$ are
	\begin{equation}
		k_a\,\frac{c}{K_d^A}
		\qquad \mbox{ and } \qquad
		k_i\,\frac{c}{K_d^I} \fs
	\end{equation}
	The rates of binding and unbinding are respectively given by
	\begin{equation}
		c\,k_b^\alpha = \frac{c}{K_d^\alpha}\,\omega_s\,e^{-\beta\varepsilon_\alpha}
		\qquad \mbox{ and } \qquad
		k_u^\alpha = \omega_s\,e^{-\beta\varepsilon_\alpha} \comma
	\end{equation}
	when it is in the activity state $\alpha$.
	The ratio between the rate constants $k_u^\alpha/k_b^\alpha$ is the \emph{dissociation constant} of the binding process, $K_d^\alpha$:
	\begin{equation}
		K_d^\alpha = \frac{k_u^\alpha}{k_b^\alpha} = c_0\,e^{-\beta(\varepsilon_b^{(\alpha)}+\mu_0)} \fs
	\end{equation}
%	The ratio $k_i/k_a$ is called \emph{allosteric constant}, $L$:
%	\begin{equation}
%		L = \frac{k_i}{k_a} = e^{\beta(\varepsilon_A - \varepsilon_I)} \fs
%	\end{equation}
	The dynamics of a single isolated protomer is depicted in Fig.\,\ref{fig:CS-single}.
	The ratios of the rate constants $k_{u,b}^\alpha$ and $k_{i,a}$ are determined by the equilibrium statistics, while their specific values affect the kinetics.
	
	It is worth noticing that the binding/unbinding rates at one protomer only depend on the state of the protomer itself and no other protomer in the ring:
	this assumption of independent binding is typical of allosteric models.
	
%	\begin{equation}\label{eq:CS-glauber-master}
%	\eqalign{
%		\frac{d}{dt} P(s,t) &= \sum_{s'} \left[ P(s',t)\,W(s'\to s) - P(s,t)\, W(s\to s')\right] \\
%		 &= (L^\dag P)(s,t) \comma
%	}
%	\end{equation}
%	where $L^\dag$ is the adjoint of the operator $L$ which defines the time evolution of averages of functions of the process $S_t$ via the equation
%	\begin{equation}
%		\frac{d}{dt} \av{A(S_t)}_s = \av{(L\,A)(S_t)}_s \comma
%	\end{equation}
%	$\av{\cdot}_s$ denoting averages taken under the condition that the process is in the state $s$ at time $t$.
	%
	%
	\begin{figure}[b]
		\centering
		\vspace*{-20pt}
		\includegraphics[scale=1]{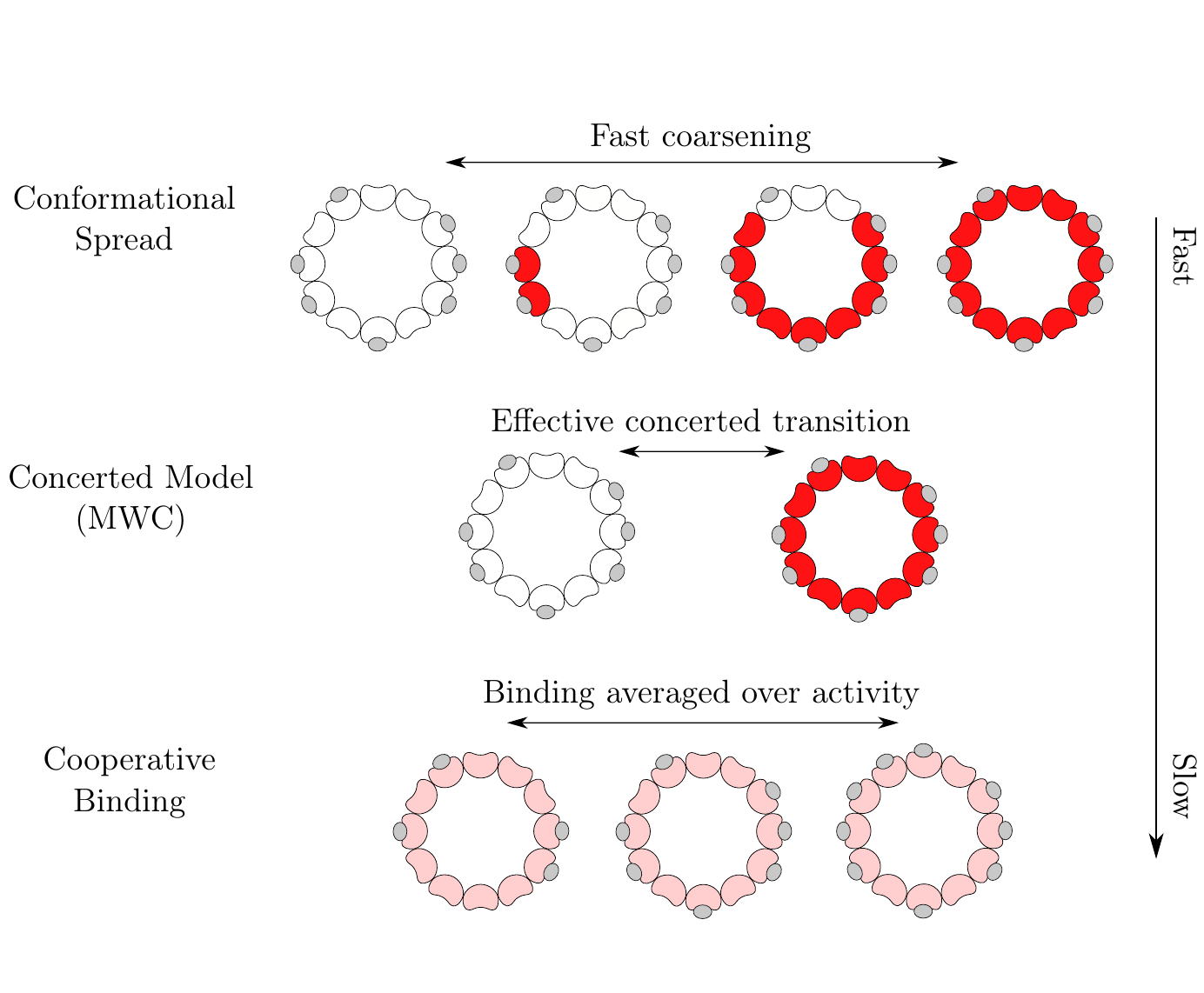}
		\vspace*{-20pt}
		\caption{\textbf{Time-scale separation in the Conformational Spread Model.}
			Graphic representation of the time-scale separation scheme.
			Short-lived transient states containing domain walls are decimated in a first time-scale separation, leading from the Conformational Spread to the MWC model, while the binding dynamics is kept frozen.
			Then, over the binding time scales the activity states are averaged out, resulting into a cooperative binding model.}\label{fig:timescales}
	\end{figure}
	Deriving an exact solution for the conditional probability $P(s,t|s_0,0)$ by directly attacking the Kolmogorov equation (\ref{eq:CS-glauber-master}) is far from being an easy task.
	However, as experiments show \cite{bai_10}, in the flagellar motor regulation mechanism of \emph{E.\,coli} it is possible to identify a hierarchy of widely separated time scales.
	This opens up the possibility of operating a reduction of the set of states by gradually \emph{integrating out}/\emph{decimating} fast degrees of freedom, operating a \emph{quasi-stationary} approximation:
	the time scale of the slow degrees of freedom is much longer than the time needed for the fast variables to relax to a stationary distribution;
	hence, the fast degrees of freedom enter the slow dynamics only through quantities averaged over such stationary distribution (conditioned to the state of the slow variables) \cite{ps_multi, weinan_multi, bo-celani}.
	The application of such techniques to the study of the allosteric regulation of the motor of \emph{E.\,coli} will be the subject of the following sections.
	The approximation scheme is depicted in Fig.\,\ref{fig:timescales}.

	\section{From the Conformational Spread to the MWC model}\label{sec:2MWC}

	In the present problem, the fastest degrees of freedom are associated with the spin-activity variables: 
	the (concerted) conformational transition between CW and CCW state is much faster than the time scale for binding/unbinding of CheY-P, respectively occurring on typical times of $10^{-3}\, s$ and $10^{-1}\, s$.
	In the associated Glauber dynamics in Eq.\,\eqref{eq:CS-glauber}, this can be encoded in the limit $\omega_f \gg \omega_b$.
%	Over timescales much shorter than the typical binding time, one can therefore think of the occupation states $\{\ell_i\}$ as parameters of a quenched external magnetic field biasing the activities $\{\sigma_i\}$.
	
	Furthermore, it can be seen that the coarsening dynamics of the spin-activity variables occurs over time scales much shorter than the typical time interval between two successive nucleations of an activity domain, the latter setting the frequency of the switch from CW to CCW and vice versa, while the binding $\{\ell_i\}$ is fixed.
	This is due to the strong coupling between the neighbouring protomers, $\beta J \gg 1$, or equivalently, $\gamma \to 1$.
	In this limit, the transition rates away from the fully aligned configurations (all $\sigma_i$ equal) are of order $\omega_f/(1-\gamma)$, while all other spin transitions are much slower, with typical rate $\omega_f \ll \omega_f/(1-\gamma)$.
	
	The discussion of this latter time-scale separation is the subject matter of this section:
	it will be shown that the strong coupling limit amounts to considering the conformational spread model effectively equivalent to the Monod--Wyman--Changeux model, on the time scale of the switch.
	At the time scales typical of these fast processes, the binding state $\{\ell_i\}$ enters via a \emph{quenched} external field term, playing a parametric role in determining the quasi-stationary distribution towards which the activity states relax.
	The slow binding dynamics will be discussed in the next section.

	\subsection*{The role of the coupling}
	
	The ferromagnetic coupling in the conformational spread model is an essential ingredient which accounts for high sensitivity of the motor to the variation of concentration of CheY-P, due to the resulting cooperative response.
	The implementation of a large coupling $J$ is suggested by the experimental determination of this high sensitivity, quantified by a Hill coefficient $\sim 10$.
	%
	%	commentare su Hill coefficient?
	%
	As pointed out in \cite{ma_12}, though, the estimation of the Hill coefficient does not impose severe constraints on the parameters of the model, especially on $J$;
	in fact, the numerical simulations performed therein show that the sensitivity depends more strongly on the activation energy of the single protomer ($\varepsilon_{A,I}$) than on the cooperativity.
	However, combining the experimental knowledge of the Hill coefficient with the information about other quantities, such as the mean locked state time and the mean switch time, Ma et al. \cite{ma_12} were able to provide a very precise estimation of $J$, which is $\sim 4.5\,k_BT$. %\footnote{CONTROLLA E VEDI DIAGRAMMA DEI LIVELLI PER BENE}.
	For such value of $J$ the formation of domain walls is strongly disfavored.
	At equilibrium, in fact, the ratio between the probability of configurations with $2m$ domains and the probability of a coherent one can be estimated as (see Ref.\,\cite{dnb_01})
	\begin{equation}
		\frac{P(2m)}{P(0)} \simeq {N \choose 2m}\,\exp(-2m \beta J) \comma
	\end{equation}
	where the binomial factor counts all possible ways of dividing $N$ protomers into $2m$ domains;
	for $N = 30 \gg 1$, the limit $P(2m)\ll P(0)$ corresponds to
	\begin{equation}
		\beta J > \log\,N \sim 3.5 = \beta J_* \comma
	\end{equation}
	satisfied by the estimate of $J$ performed in \cite{ma_12}.
	The stationary equilibrium configuration, at fixed binding states $\{\ell_i\}$, is therefore concentrated only on the two states with all the protomers in the same state.
	From a dynamical point of view, this means that states with one or several domain walls are just short-lived transients between coherent states:
	as soon as a domain is nucleated inside a coherent configuration, it either immediately expands to invade the whole ring or is suddenly absorbed, typically much before another nucleation occurs.

	\subsection*{Decimation of fast variables}

	To realize the fast ``emptying'' of configurations with several domain walls, it is necessary to analyse the structure of the transition rate matrix $K(s\to s')$, when the limits of the time-scale separation ($\omega_s \ll \omega_f \ll \omega_f/(1-\gamma)$) are concerned. 
	
\begin{figure}[t]
		\centering
%		\small\def\svgwidth{.75\textwidth}
%		\input{emptying_plain.pdf_tex}
		\includegraphics[width=.75\textwidth]{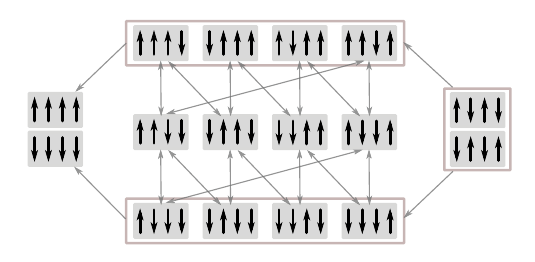}
		\caption{\textbf{Fast coarsening dynamics.}
		Schematic representation of the fast rates of single-spin flipping $K_f$ for a system of 4 protomers.
		Grey boxes correspond to the states; periodic boundary conditions are understood.
		%The rates $K^{(i)}_f$ rule the coarsening of the spin-activity variables.
		Arrows are drawn between two states (or groups of states) for which $K_f$ is non vanishing for some $i$. In particular:
		reversible transitions are allowed between states with equal number of domain walls;
		transitions to states with less domain walls are irreversible.
		Starting from any state, the dynamics leads to one of the coherent configurations in a time $\sim (1-\gamma)\,{\omega_f}^{-1}$ [see Eqs.~\eqref{eq:CS-glauber} and \eqref{eq:glauber-fast}];
		such states are the only two activity states in the MWC allosteric model.
		}\label{fig:coarsening}
	\end{figure}

	In the limit $\gamma \to 1$, in fact, the non-vanishing entries of the matrix $K$, at frozen binding $\{\ell_i\}$, are either of order $\omega_f/(1-\gamma)$, or of order $\omega_f$:
	the latter rates (slow) are defined for transitions consisting in a nucleation of a domain, \emph{i.e.} creation of pairs of domain walls, and are denoded by $K_s$;
	the former (fast) are defined for all other transitions, \emph{i.e.} motion and destruction of domain walls, and are denoted by $K_f$.
	We can therefore write $K = K_f + K_s$, with
	\begin{equation}\label{eq:glauber-fast}
		K_f(s\to s') = K(s\to s')\,(1 - \delta_{\sigma_1\ldots\,\sigma_N}) \sim \frac{\omega_f}{1-\gamma} \comma
	\end{equation}
	and
	\begin{equation}\label{eq:glauber-slow}
		K_s(s\to s') = K(s\to s')\,\delta_{\sigma_1\ldots\,\sigma_N} \sim \omega_f \comma
	\end{equation}
	where $K$ are defined in Eq.\,(\ref{eq:CS-glauber}), and $\delta_{\sigma_1\ldots\,\sigma_N}$ indicates that the spin-activity variables in $s$ have all the same value.
	One notices that the coherent configurations (all protomers active or inactive) are the only absorbing states of the fast process, since in such cases the entries of $K_f$ vanish.
	The dynamics specified by $K_f$ forbids the creation of pairs of domain walls and only allows translation or absorption of domain walls.
	As a result, the fast dynamics leads to one or the other coherent configuration with a typical rate $\sim \omega_f/(1-\gamma)$.
	As an explicative example, the case of $N = 4$ is depicted in Fig.\,\ref{fig:coarsening}.
	On a time scale set by $1/\omega_f$, the nucleation of an activity domain can occur.
	In the coherent activity configurations, the process involving the spin-activity variables has slow rates $K_s$.
	It is then possible to apply the standard techniques of time-scale separation \cite{ps_multi, weinan_multi, bo-celani}, eliminating incoherent activity configurations from the dynamics at time scales comparable with $1/\omega_f$ or longer.
	The net effect of the fast coarsening dynamics is included in an effective way into rates, denoted by $K_c$, which provide the description of the dynamics at the nucleation time scale:
	a concerted transition between the two coherent configurations $I$ (all protomers inactive, $\sigma_i=-1$) and $A$ (all active, $\sigma_i = 1$), besides slow binding processes.
	In this model, the $N$-protomer complex can be in 2 different activity states, each of which present in $2^N$ binding configurations (2 for each protomer):
	therefore, the model contains $2\times 2^N$ states, and corresponds to the concerted allosteric model of Monod, Wyman and Changeux (MWC) \cite{mwc_65, mgp_13}.
	
\begin{figure}[t]
		\centering
%		\def\svgwidth{.7\textwidth}
%		\small\input{immagini/MWCgeneralized.pdf_tex}
		\includegraphics[scale=1]{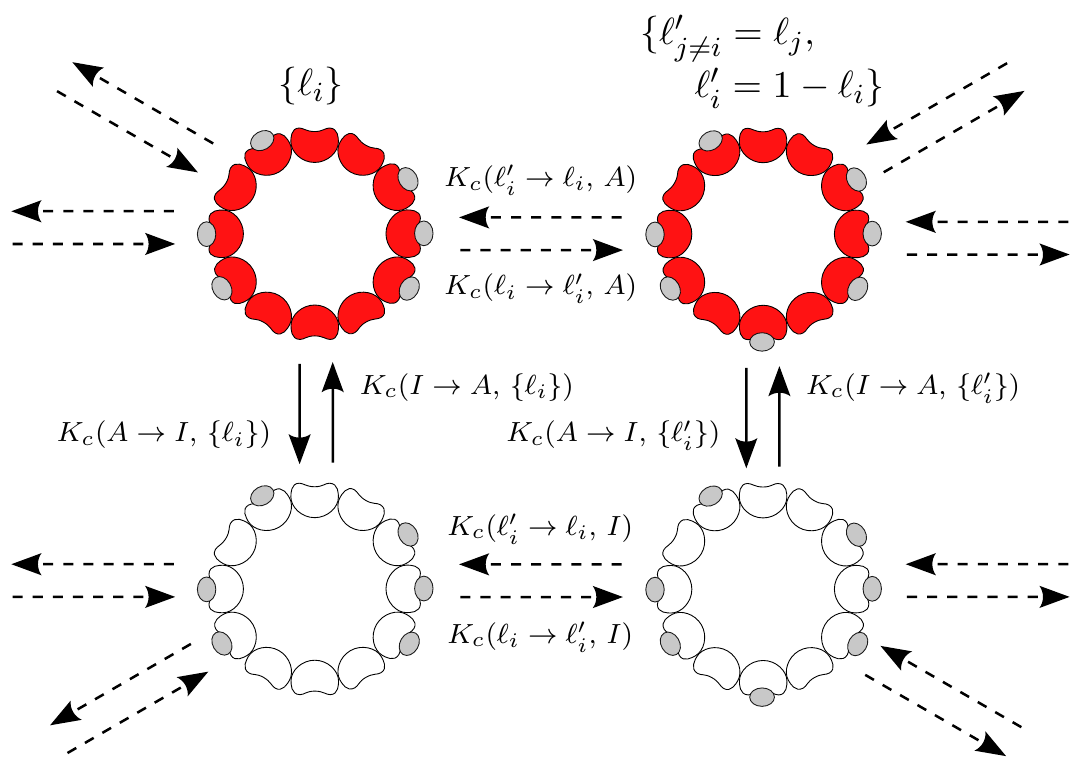}
		\caption{\textbf{Dynamics of the MWC model.}
			The figure contains only a small portion of the model, corresponding to two possible binding states, $\{\ell_i\}$ and $\{\ell'_i\}$ (differing only by the occupation of the protomer at the bottom of the ring).
			The spin-activity variables of the ring of protomers are involved in a fast concerted transition (solid arrows), with rates $K_c$ depending on the occupation $\{\ell_i\}$ in a highly non trivial way. 
			The transitions between different binding configuration is slow (dashed arrows); only one of the possible binding/unbinding transitions is explicitly represented, while the others are symbolically indicated by unlabeled arrows. 
		}
		\label{fig:MWCgen}
	\end{figure}

	The structure of the state diagram of the MWC model with its rates $K_c$ is depicted in Fig.\,\ref{fig:MWCgen}.
	In the Appendix, the decimation procedure leading from the conformational spread to the MWC model, in the case of $N=2$ has been worked out exactly.
	In general, the rate of a concerted switch from the activity state $\alpha=I$ (or $A$) to $\alpha'= A$ (or $I$) is
	\begin{equation}\label{eq:eff-MWC}
		K_c(\alpha \to \alpha',\,\{\ell_i\}) = \sum_{j=1}^N K_s (\alpha \to \alpha^{(j)},\,\{\ell_i\}) \, P_{abs}^{(j)}(\alpha') \comma
	\end{equation}
	where $\alpha^{(j)}$ denotes the state where all the spins but the $j$-th are in the state $\alpha$, and $P_{abs}^{(j)}(\alpha')$ is the probability of absorption in the state $\alpha'$ conditioned to the initial state $\alpha^{(j)}$.
	
	A direct analytic derivation of the rates $K_c$ (or the probabilities $P_{abs}^{(j)}$) for a generic $N$-protomer ring can be extremely complicated.
	However, in the time-scale separation assumptions, the fast dynamics after the nucleation of an activity domain from a coherent state reaches one of its 2 absorbing states before another nucleation could possibly occur.
	This means that a calculation of the effective activity switching rates in the MWC model, does not require to include all the incoherent states, but only those with just two domain walls:
	the coarsening process can be seen as the expansion or contraction of the domain which has been nucleated.
	The nucleated domain can either expand until it invades the whole ring (complete switch), or be ``absorbed'' back (failed attempts).
	
	Since the detailed balance is still respected by the rates in the decimated dynamics, all their pairwise ratios are fixed by the equilibrium distribution.
	Hence, since the equilibrium distribution of the MWC model is known from the Hamiltonian \eqref{eq:CS-ham} (where the coupling part is just a constant term), it is sufficient to determine only \emph{one} effective rate $K_c$ exactly.
	In the case where $\ell_i = 0$ for all protomers, one is able to calculate the rate of switching from the $I$ to the $A$ state, by mapping the coarsening process into a simple birth and death process, the random variable being the size of the domain with active protomers (see the Appendix).

	Regarding the binding process, the rate $K_c$ is just the binding/unbinding contribution in the rates $K$, defined in Eq.\,(\ref{eq:CS-glauber}).
	
	Although the dynamics of the MWC model depends on the detailed binding configuration $\{\ell_i\}$ in a highly non-trivial way, the equilibrium distribution depends on the total occupancy $l = \sum \ell_i$ only,
	\begin{eqnarray}
		P_{eq}(I,\,l) &= \frac{\left(\frac{c}{K_d^I}\right)^l\,{N\choose l}}{\left(1 + \frac{c}{K_d^I}\right)^N + L^{-1}\left(1 + \frac{c}{K_d^A}\right)^N} \comma \label{eq:eq_Al} \\[1ex]
		P_{eq}(A,\,l) &= \frac{L^{-1}\left(\frac{c}{K_d^A}\right)^l\,{N\choose l}}{\left(1 + \frac{c}{K_d^I}\right)^N + L^{-1}\left(1 + \frac{c}{K_d^A}\right)^N} \label{eq:eq_Il} \comma
	\end{eqnarray}
	where $L$ is called \emph{allosteric constant} of the $N$-protomer MWC molecule,
	\begin{equation}
		L = \bigg(\frac{k_i}{k_a}\bigg)^N = e^{\beta\,(\varepsilon_A - \varepsilon_I)\,N} \fs
	\end{equation}

	There is an important comment to be made about the equilibrium distribution of the MWC model, in particular about the marginal probability for the active state, defined as the \emph{activity} of the MWC molecule,
	\begin{eqnarray}\label{eq:n-act}
		P_{eq}(A) &= \sum_{l=0}^N P_{eq}(l,\,A) = \frac{1}{1 + L\left(\frac{K_d^A}{K_d^I}\right)^N\left(\frac{c+K_d^I}{c + K_d^A}\right)^N} \label{eq:n-eq-act}\fs
	\end{eqnarray}
	In our problem, this corresponds to the CW \emph{bias} of the flagellar motor, which is a function of the CheY-P concentration $c$.
	In order for the MWC molecule to be a good allosteric switch, it needs to be almost certainly active for high enough concentration $c$ and, vice versa, inactive when $c$ is low:
	\[
		P_{eq}(A) \sim \left\{
		\eqalign{
			&\left[1 + L\right]^{-1} \to 0  \quad \mbox{for \ } c\to 0 \\
			&\left[1 + L^{-1}\left(\frac{K_d^A}{K_d^I}\right)^N\right]^{-1} \to 1 \quad \mbox{for \ } c\to \infty
		}
		\right. \fs
	\]	
	These limits impose the following constraints:
	\begin{equation}\label{eq:act-constr}
		1 \ll L \ll \left(\frac{K_d^I}{K_d^A}\right)^N \fs
	\end{equation}
	\begin{figure}[b]
		\centering
		\includegraphics[width=.5\textwidth]{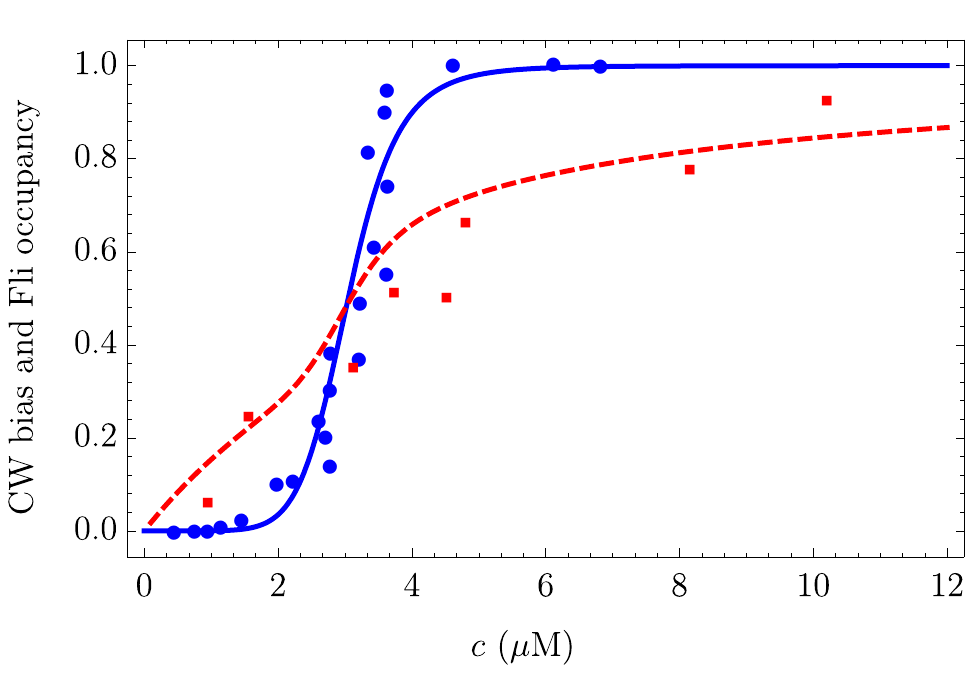}
		\caption{\textbf{Activity and mean Fli occupancy at equilibrium.}
			Analytic results for $P_{eq}(A)$ (solid blue line) and mean Fli relative occupancy $\av{l}/N$ (dashed red line) as a function of the CheY-P concentration, $c$. The dots are the experimental results presented in Ref.\cite{berg_03}.
			In our work we chose the dissociation constants to be $K_d^A = 1.84\,\mu M$ and $K_d^I = 5.52\,\mu M$, respectively, while the allosteric constant has been set to be $L =10^{7}$.
			The plot shows the effect of allostery:
			the activity response is much more sensitive than the binding to changes of concentration of CheY-P.
		}\label{fig:act_vs_c}
	\end{figure}
	Since the single protomer has higher ligand affinity (smaller dissociation constant $K_d$) when in the active state than in the inactive one, it is required that $K_d^A < K_d^I$.
	From this last relation one realizes that the number of protomers sets the \emph{sensitivity} of the switch:
	since $K_d^I > K_d^A$, the larger $N$, the larger the r.h.s of the condition given by Eq.~\eqref{eq:act-constr}.		
	Incidentally, depending on environmental stimuli \emph{E.\,coli} is able to regulate the number of protomers of the flagellar motor \cite{delalez_10, lele_12, berg_12}.
%	one may ask whether the adaptation of the number of protomers  the optimal sensitivity that \emph{E.\,coli} needs in a certain environment.

	\section{From MWC to a cooperative binding model}\label{sec:2BD}
	
	As we already said at the beginning of Sec.\,\ref{sec:2MWC}, the binding is much slower than the switching dynamics.
	We can assume that on the time scale at which one of the protomers binds or releases a CheY-P (set by a typical time $\tau_b\sim 10^{-1}\,s$), the activity of the ring safely reaches the equilibrium configuration, conditioned to the (quasi-static) value of $l$:
	\begin{eqnarray}
		P_{eq}(I|l) &= \frac{P_{eq}(l,I)}{P_{eq}(l)} = \frac{P_{eq}(l,I)}{P_{eq}(l,I) + P_{eq}(l,A)} = \frac{1}{1 + L^{-1}\left(\frac{K_d^I}{K_d^A}\right)^l} \comma \label{eq:inact-given-l} \\[1ex]
		P_{eq}(A|l) &= \frac{P_{eq}(l,A)}{P_{eq}(l)} = \frac{P_{eq}(l,A)}{P_{eq}(l,I) + P_{eq}(l,A)} = \frac{1}{1 + L\left(\frac{K_d^A}{K_d^I}\right)^l} \label{eq:act-given-l} \fs
	\end{eqnarray}
	Then, on time scales comparable to (or larger than) $\tau_b$, the relevant dynamics is essentially the slow binding/unbinding one, while the fast activation/inactivation dynamics is \emph{averaged} over the equilibrium conditional probabilities in Eqs.\,\eqref{eq:inact-given-l} and \eqref{eq:act-given-l}, to give the effective rates $\bar{K}$ for the variable $l$:
	\begin{equation}\label{eq:eff-rates}
			\bar{K}(l \to l') = \sum_{\alpha\in\{I,A\}} P_{eq}(\alpha|l)\,K(l\to l',\,\alpha\to \alpha) \fs
	\end{equation}
	This averaging procedure is guaranteed to give an effective dynamics of the slow variables which still enjoys the Markov property.
	The effective binding/unbinding rates of the whole allosteric complex are, in fact,
	\begin{equation}\label{eq:eff-rates2}
		\eqalign{
		\bar{K}(l \to l+1) &= (N-l)\,c\,\bar{k}_b^{(l)} \equiv b_l \comma \\
		\bar{K}(l \to l-1) &= l\,\bar{k}_u^{(l)} \equiv u_l \comma
		}
	\end{equation}	
	where
	\begin{equation}
		\bar{k}_{b,u}^{(l)} = \frac{k_{b,u}^A}{1 + L\left(\frac{K_d^A}{K_d^I}\right)^l} + \frac{k_{b,u}^I}{1 + L^{-1}\left(\frac{K_d^I}{K_d^A}\right)^l} \comma
	\end{equation}
	depending only on the current  value of $l$.

	A comment about the range of validity of this result is in order: for the time-scale separation to hold, the rates $\bar{K}$ must be small enough to guarantee that the binding/unbinding process is still much slower than the activation/inactivation.
	In particular, this implies that the concentration of ligands in the environment $c$ cannot be exceedingly large;
	then, in the time-scale separation approximation,  we keep ourselves far from this regime.
	
	The reduced system is also a Markov process, governed by the following master equation:
	\begin{equation}\label{eq:eff-mwc-master}
%		\partial_t P_t(l) = (N-l+1)\,c\,\bar{k}_b^{(l-1)}\,P_t(l-1) + (l+1)\,\bar{k}_u^{(l+1)}\,P_t(l+1) \\ - \left[ l\,\bar{k}_u^{(l)} + (N-l)\,c\,\bar{k}_b^{(l)} \right]\,P_t(l)
		\partial_t P_t(l) = b_{l-1}\,P_t(l-1) + u_{l+1}\,P_t(l+1) - \left[ b_{l} + u_{l} \right]\,P_t(l) \fs
	\end{equation}
	The process hence obtained is a \emph{birth-and-death process}, restricted on the set of integers between $l=0$ and $l=N$.
	These extremes are reflecting boundary states.
	This dynamics eventually leads to the equilibrium state $P_{eq}(l)$, easily calculated by marginalizing the joint probability distribution $P_{eq}(\alpha,\,l)$,  given in Eqs.\,\eqref{eq:eq_Al} and \eqref{eq:eq_Il}:
	\begin{equation}\label{eq:eq_l}
		P_{eq}(l) = P_{eq}(A,\,l) + P_{eq}(I,\,l) \fs
	\end{equation}

	Albeit much reduced, this model still encodes a lot of information about the actual dynamics of the switch.
	Indeed, the flagellar motor switch is triggered by the number of ligands bound to the allosteric complex.
	In the next section we present some numerical analysis of the dynamical properties of the effective cooperative binding model obtained above.

	\section{Dynamics of the effective cooperative binding model}\label{sec:BD}
	
	In this Section we analyze the case of a motor constituted by $N=30$ Fli molecules.
	The allosteric constant $L$ and the dissociation constants $K_d^A$ and $K_d^I$ have been chosen consistently with Ref.\,\cite{berg_03} and works cited therein: $L = 10^{7}$, $K_d^A = 1.84\; \mu M$ and $K_d^I = 5.52\;\mu M$;
	these values provide a qualitatively good fit of the activity as a function of the CheY-P concentration $c$ (see Fig.\,\ref{fig:act_vs_c}).
%	The recommended value for the allosteric constant $L$ is $\sim 10^7$:	
%	The very large value of $L$ means that, for an unbound protomer, the inactivation process is much faster than the activation one
%	The ones defined so far, together with the concentration of CheY-P, are the only quantities which uniquely define the equilibrium properties of the MWC model for our motor.
	With this choice of the parameters, we can easily see that the bound in Eq.\,\eqref{eq:act-constr} is safely satisfied, so that the motor displays a switch behaviour, manifest in the response curve in Fig.\,\ref{fig:act_vs_c}.
	One also notices that the motor \emph{operates} within a range of concentration $c$ roughly between $K_d^A$ and $K_d^I$.
	The maximum sensitivity is found around a value $c_*$, which correspond to a CheY-P concentration such that the CW (active) and the CCW (inactive) states occur with equal probabilities at equilibrium.
	
	As already remarked above, the specific values of the rate constants $k_{b,u}^\alpha$ are irrelevant for the equilibrium properties of the model, but they determine the characteristic time scale for the motor switch.
	Out of these four constants, only two are actually independent, since we already defined their ratios $K_d^A = k_u^A/k_b^A$ and analogously $K_d^I = k_u^I/k_b^I$.
	Then, the dynamics of the cooperative binding model can be specified only the parameters $k_b^I$ and $k_b^A$;
	the qualitative behaviour is determined only by their ratio, while their specific values gives information about the overall time scale (of the binding process).
	In our work, we set $k_b^A = 2.8\,s^{-1}$ and $k_b^I = 5.0\,s^{-1}$, consistently with those recommended by Bai et al. \cite{bai_10}.
	\begin{figure}[t]
		\centering
		\includegraphics[width=.45\textwidth]{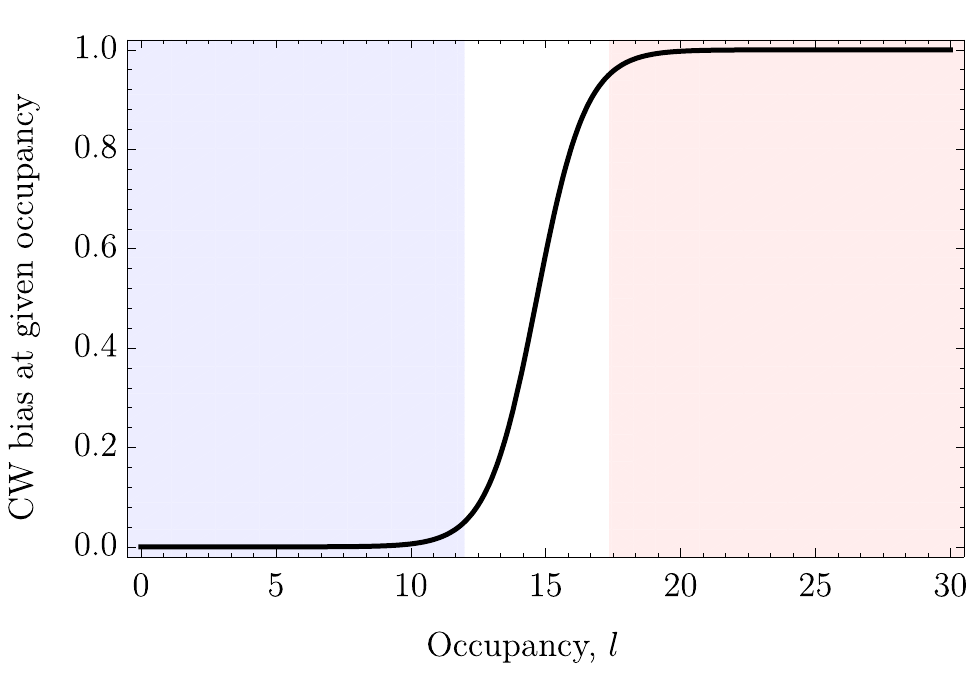}
		\hspace*{.03\textwidth}
		\includegraphics[width=.45\textwidth]{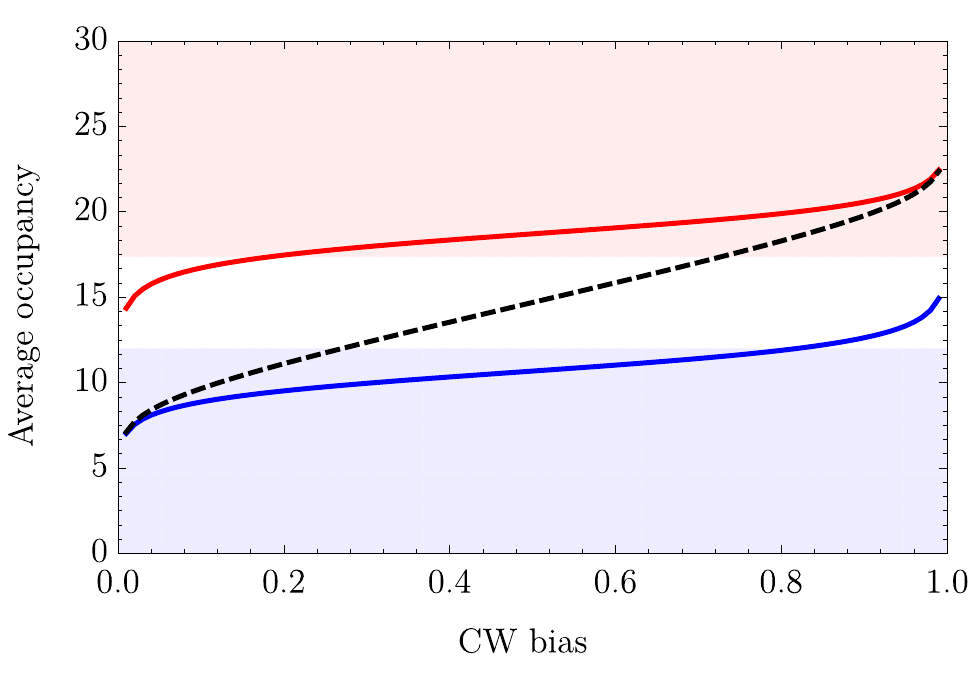}
		\caption{
		\textbf{Motor switch ruled by the cooperative binding.}
		On the left,
		the probability of the CW state conditioned on the Fli occupancy; for values of $l$ fixed in the shaded regions, the motor is in the CW or CCW state with 95\% probability.
		On the right,
		average occupancy as a function of the CW bias: the dashed line corresponds to the unconditional average (see also Fig.\,\ref{fig:act_vs_c}) while the solid lines represent the averages conditioned to the CW state (red), $\bar{l}_A$, and CCW state (blue), $\bar{l}_I$.
		The values of $\bar{l}_A$ and $\bar{l}_I$ lie in the respective 95\%-confidence intervals, with a CW bias between $\simeq 0.1$ and $\simeq 0.9$.
		The locked--state time can be interpreted as the first passage time between $\bar{l}_A$ and $\bar{l}_I$ in the cooperative binding model.
		}\label{fig:cond_l}
	\end{figure}

	As previously discussed, the cooperative binding model obtained so far must provide an accurate description of the statistics of slow observables, namely those which vary over time scales typical of the binding process or longer.
	From experimental results, it is clear that the mean--locked state time (\ie the time in which the motor stays in a certain rotational state between two consecutive switches) is such an observable;
	we show, indeed, that the cooperative binding model captures very well its statistics.
	
	Let us denote by $\bar{l}_I$ and $\bar{l}_A$ the averages of the occupancy $l$ conditioned, respectively, to the inactive state (CCW) and active state (CW).
	%Notice that $\bar{l}_I$ is smaller than the unconditional average $\langle l \rangle = \bar{l}_I P_{eq}(I) + \bar{l}_A P_{eq}(A)$, while $\bar{l}_A$ is larger (see Fig.\,\ref{fig:cond_l}, right).
	One can see that the probability of the CW state is very close to unity if the Fli occupancy is conditioned to $\bar{l}_A$, and almost vanishing when conditioned to $\bar{l}_I$ (Fig.\,\ref{fig:cond_l}).
	Therefore, since the fast activity variables are slaved to the slow binding ones, we can state that a good measure of the locked--state time  is the first passage time between $\bar{l}_A$ and $\bar{l}_I$.

	Let us then study the first arrival time at $\bar{l}$ from a generic state $k$.
	If we denote by $f_k$ the probability density function of this time interval, its moment generating function
	\begin{equation}\label{eq:genfFPT}
		g_k(\lambda) = \int_0^\infty d\tau\, e^{-\lambda\,\tau}\,f_k(\tau) \comma
	\end{equation}
	satisfies
	\begin{equation}\label{eq:genfFPT-2}
		\sum_l g_l(\lambda)\,(M_{l,k} - \lambda\,\delta_{l,k}) = - \delta_{k,\,\bar{l}} \comma
	\end{equation}
	where  \textbf{M} is the generator of the process in which absorbing conditions have been put at $\bar{l}$.
	From Eq.\,\eqref{eq:genfFPT-2}, we can derive the equation for the mean first passage time at $\bar{l}$, using $\av{\tau_k} = g'_k(\lambda)|_{\lambda=0}$:
	\begin{equation}\label{eq:meanFPT}
		\sum_l \av{\tau_l}\,M_{l,k} = -1 \fs
	\end{equation}
	\begin{figure}[t]
		\centering
		\includegraphics[width=.45\textwidth]{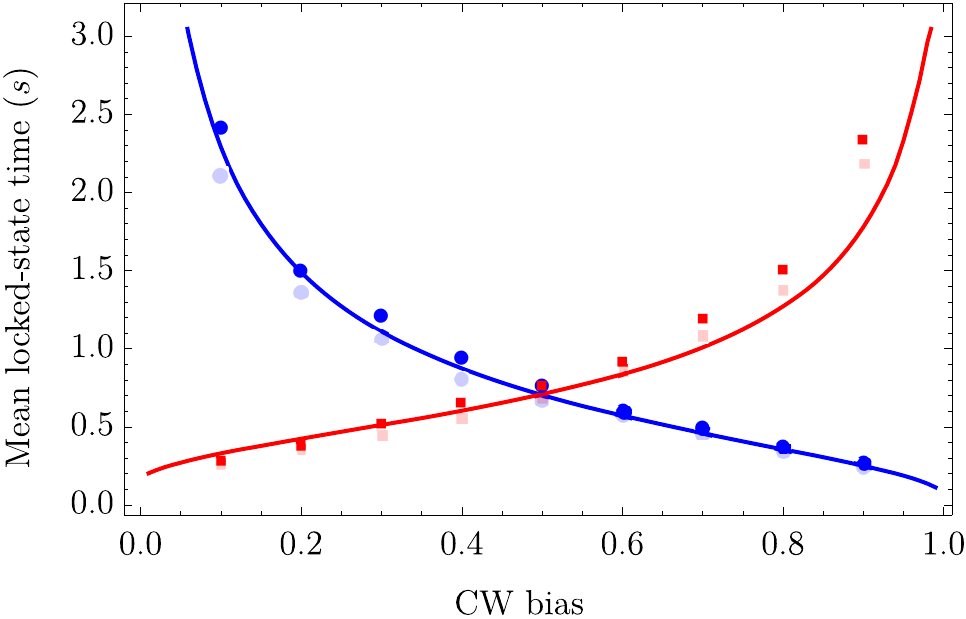}
		\hspace*{.03\textwidth}
		\includegraphics[width=.45\textwidth]{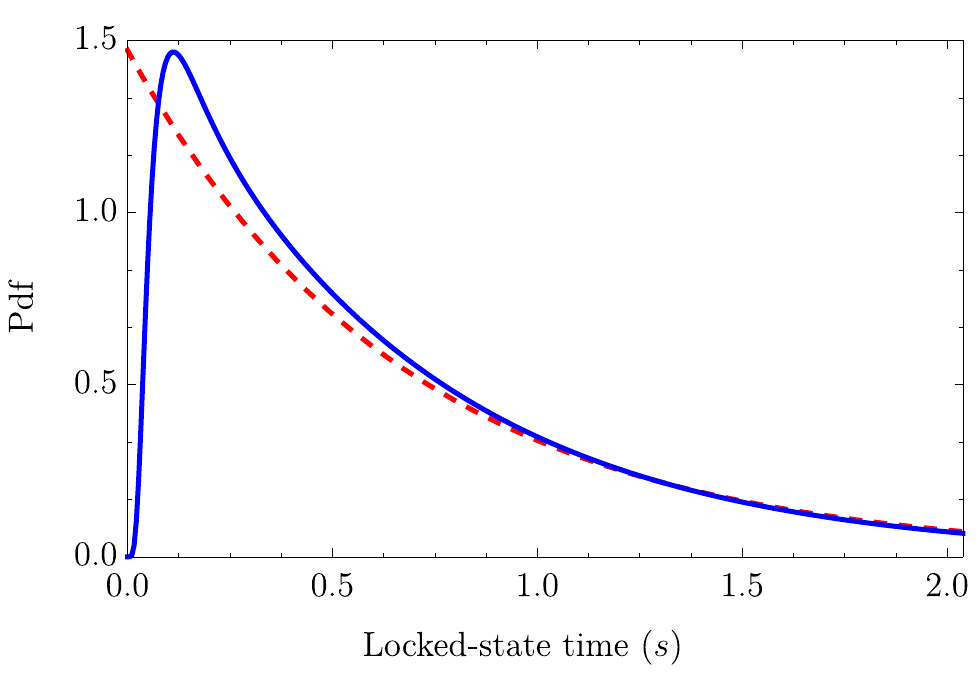}
		\caption{%
		\textbf{Statistics of the locked--state time.} On the left, average locked--state time in the CCW state (blue line and circles) and CW state (red line and squares): the points are experimental (lighter color) and numerical (darker color) results from \cite{bai_10}; the lines are the theoretical results from the cooperative binding model, estimated as the first passage time at $\bar{l}_a$ conditioned to $\bar{l}_i$ at $t=0$ (mean CCW time, blue), and vice versa (mean CW time, red).
		On the right, probability distribution of the first passage time at $\bar{l}_a$ from $\bar{l}_i$ in the unbiased case:
		the solid blue line is the exact result found as the inverse Laplace transform of the generating function obtained by solving Eq.\,\eqref{eq:genfFPT-2}; the dashed red line is the exponential distribution with the same average. See for comparison the experimental and numerical results in Refs.\cite{bai_10,ma_12}.}\label{fig:FPT}
	\end{figure}
	Exact results are obtained by inverting Eq.\,\eqref{eq:meanFPT} and are shown in Fig.\,\ref{fig:FPT}, with $k=\bar{l}_i$ and $\bar{l} = \bar{l}_a$, and vice versa, for several values of the bias $P_{eq}(A)$.
%	We notice that the mean first passage time is well captured by the Langevin dynamics with constant diffusion coefficient discussed above;
%	the result is closer to the purely diffusive one as the initial state $k$ approaches $\bar{l}_a$, because of the very small drift.

	We also extract the probability density by solving Eq.\,\eqref{eq:genfFPT-2} and numerically performing the inverse Laplace transform.
	The resulting distribution is  very similar to the experimental and numerical results presented in \cite{bai_10} and \cite{ma_12}, confirming that the effective cooperative binding model gives an excellent description of the motor kinetics.

	\section{Discussion}
	
	In this work we pursued an analytic approach to the description of the dynamics of the conformational spread model, a phenomenological model which well reproduces the allosteric regulation of the flagellar motor in \emph{E.\,coli}.
	Our analysis was based on the existence of a hierarchy of widely separated time scales in the biochemistry of the motor of \emph{E.\,coli}. Namely, over the scale of conformational transitions between CW and CCW states in the Fli molecules (protomers, constituents of the flagellar motor) incoherent states are very short-lived, and only coherent states of activity are sufficiently long-lived.	In such a limit we have reduced the conformational spread to the well known Monod--Wyman--Changeux model. For a motor with $N=30$ protomers, this approximations amounts to reducing the number of states in the model from $4^N \sim 10^{18}$ to $2(N+1)=62$.
	
	Moreover, the binding of CheY-P to the Fli molecules occurs much less frequently than the switch from a completely active to inactive state allowing to average out the fast activity states under quasi-stationary Fli occupancy (number of CheY-P bound to the motor).
	This allowed to reduce the number of states further and get a cooperative binding model containing only $N+1=31$ states, the possible values of the overall occupancy.
	The resulting Markov process is a birth-and-death process which can be studied semi-analytically, with virtually no computational cost.
	
	This effective model for the slow variables is able to capture the dynamics of observables varying on time scales of $10^{-1}\,s$ or longer.	Two of such observables are the CW and CCW locked-state time, which correspond to the duration of tumbles and runs, respectively, with time scales typically of the order of seconds.
	We showed that our model reproduces the statistics of the locked state time and is in extremely good quantitative agreement with experimental measurements.
	
	In perspective, our approach could be extended to include the even slower kinetics of motor remodeling. Indeed, it is known that over time scales much longer than the binding times (typically minutes), \emph{E.\,coli} is also able to modify the flagellar motors by changing the number of Fli molecules, \emph{i.e.} the protomers \cite{delalez_10, lele_12}. This mechanism provides an adaptation layer at the output and restores the sensitivity of the motor when  CheY-P concentration are kept off the dynamic range for a long time \cite{berg_12}.
	
		Finally, experimental work on flagellar motors in Vibrio alginolyticus \cite{xie_15} has shown a nontrivial locked-state time-statistics.  The techniques exploited in our work might prove useful to address theoretically the origin of these observations.

	\section*{Acknowledgements}
	
	We are grateful to Stefano Bo for illuminating insights and discussions.

	\vspace*{4ex}	
	
	\setcounter{section}{0}
	\setcounter{equation}{0}
	\setcounter{figure}{0}
	\renewcommand{\thesection}{\Alph{section}}
	\renewcommand{\theequation}{\thesection.\arabic{equation}}
	\renewcommand{\thefigure}{\thesection\arabic{figure}}
	\titleformat{\section}{\normalfont\bfseries}{Appendix.}{0.5em}{}

	\section{Decimation of the fast coarsening dynamics\\ in the Conformational Spread}\label{app:decimation}

	In this appendix we show that the decimation of the short-living incoherent states in the Glauber-like dynamics of conformational spread model leads to the MWC model, in which the rates generally depend on the full binding state $\{\ell_i\}_{i=1}^N$ (kept frozen at this step).
	Nevertheless, the equilibrium properties of the resulting Markovian model only depend on the global variable $l=\sum \ell_i$.
	
	For the sake of simplicity, we will describe in detail the case of $N=2$.
	In the case of generic $N$, and in the limit of large coupling, we compute exactly the probability of completion of the switch after the nucleation of one domain, when all the protomers are unbound ($\ell_i = 0$):
	this provides information about the overall time scale of the concerted switching process;
	the switching rates in presence of a generic occupation can be then obtained from this by means of the detailed balance condition.

	\subsection*{Coarsening with 2 protomers}

\begin{figure}[b]
		\centering
		\vspace*{-3ex}
		\includegraphics[scale=1.2]{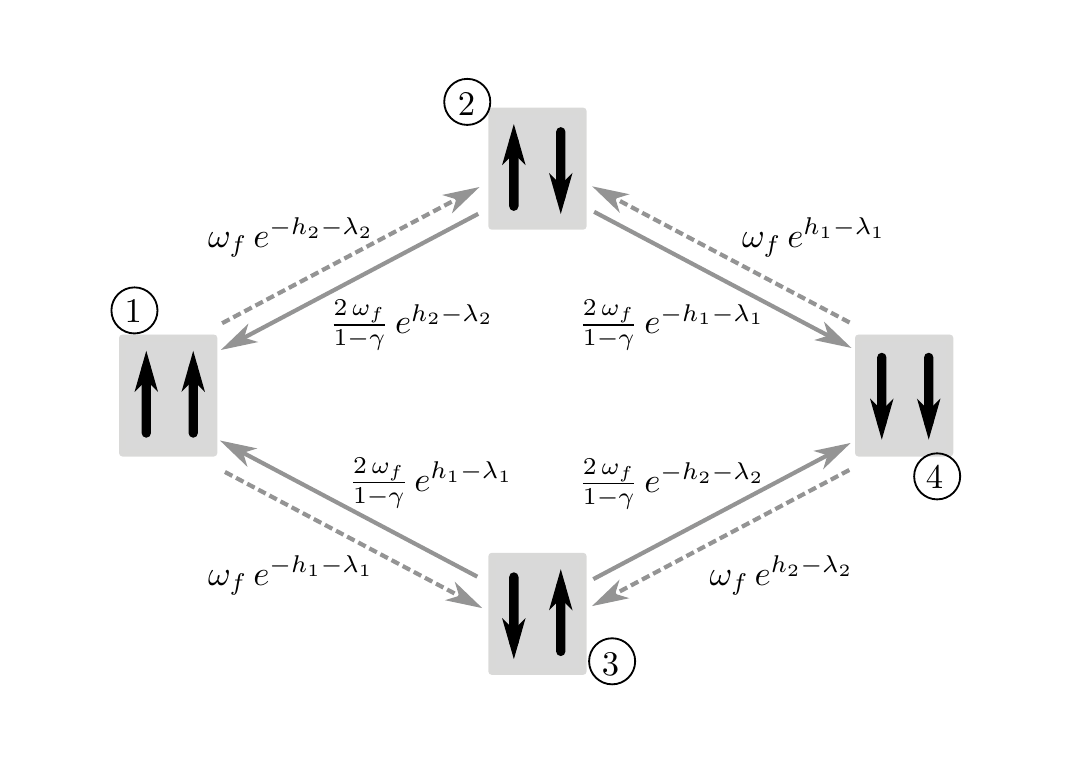}
		\vspace*{-3ex}
		\caption{\textbf{Time-scale separation  in the dynamics of the spin-activity variable.}
			Solid and dashed arrows respectively indicate fast and slow transition rates.
			Numbers within circles are labelling of the states.
			Configurations with anti-parallel spins are quickly emptied, and coherent configurations are rarely escaped (they are absorbing state for the fast dynamics). 
			In the case of $N=2$ protomers reproduced here, there are two paths joining one coherent configuration and the other.
		}\label{fig:coarse2}
	\end{figure}

	At frozen binding, the dynamics of the activity variables is equivalent to the one of $N$ spins with nearest-neighbour ferromagnetic interaction with strength $J$, subjected to a ``magnetic field'' given by Eq.~\eqref{eq:CS-energies}:
	\begin{equation}\label{eq:magfield}
		\beta\,h(\sigma_i,\,\ell_i) = h_i\,\sigma - \lambda_i \comma
	\end{equation}
	where
	\begin{equation}\label{eq:magfield-2}
		h_i = \frac{\beta}{2}\bigg[\varepsilon_I - \varepsilon_A - (\varepsilon_b^{(I)} - \varepsilon_b^{(A)})\ell_i \bigg] \comma\quad
%		\qquad\mbox{and}\qquad
		\lambda_i = \frac{\beta}{2}\bigg[\varepsilon_I + \varepsilon_A - (\varepsilon_b^A + \varepsilon_b^I + 2\mu )\ell_i \bigg] \fs
	\end{equation}
	According to Eq.\,(\ref{eq:CS-glauber}), and to the considerations given in Sec.\,\ref{sec:2MWC}, neglecting the binding dynamics, we can decompose the full matrix of the rates $K$, when $\gamma \to 1$, as the sum the fast contribution $K_f$,
	\begin{equation}
		K_f = \omega_f\,\frac{1+\gamma}{1-\gamma}
			\left(
				\begin{array}{cccc}
					0 &  e^{h_2-\lambda_2} &  e^{h_1-\lambda_1} & 0 \\
					0 & 0 & 0 & 0 \\
					0 & 0 & 0 & 0 \\
					0 & e^{-h_1-\lambda_1} & e^{-h_2-\lambda_2} & 0
				\end{array}
			\right) \comma
	\end{equation}
	and a slow part $K_s$,
	\begin{equation}
		K_s = \omega_f
		\left(
			\begin{array}{cccc}
				0 & 0 & 0 & 0 \\
				e^{-h_2-\lambda_2} & 0 & 0 & e^{h_1-\lambda_1} \\
				e^{-h_1-\lambda_1} & 0 & 0 & e^{h_2-\lambda_2} \\
				0 & 0 & 0 & 0 \\
			\end{array}
		\right) \comma
	\end{equation}
	where the row and the column index respectively correspond to the final and initial state, labelled as in Fig.\,\ref{fig:coarse2}.
	We notice that the fast dynamics has two absorbing states, which are the equilibrium configurations in the time-scale separation limit:
	these states are the coherent configurations, namely those with all the protomers in the same activity state (all spins aligned).
	
	On the slow time scales (much longer than the coarsening process but much shorter than the binding), we can calculate effective rates of passing from one coherent state to the other, given by Eq.\,(\ref{eq:eff-MWC}).
	These rates are limited by the rate of flipping one spin from the starting coherent configuration: this is the slow process, since such transition costs an energy $\sim J \gg \beta$.
	The rates are then affected by the probabilities that, once this flip has occurred, the process reaches the other coherent state and is not absorbed back in the starting coherent state:
	such probabilities are completely determined by the fast dynamics.	
	The difficulty in calculating such probability stems from the fact that, for a generic number of protomers $N$, a huge number of paths contributes, with amplitudes strongly dependent on the binding configuration $\{\ell_i\}$.
	The general way of proceeding is presented in Refs.\cite{ps_multi,weinan_multi, bo-celani}.
	
	In the simple example where $N=2$, there are only two paths which give contribution to the concerted transition:
	the flip of the first spin followed by the flip of the second one, or the flip of the second followed by the flip of the first one.
	Summing up the rates of these possible channels yields the effective rates of the concerted switch~\footnote{The parametric dependence of the rates on the binding state $\{\ell_i\}$ is understood, but not made explicit in the notation.}:
	\begin{equation}
	\eqalign{
		K_c (I\to A) = K_s&(\downarrow\!\downarrow \to \uparrow\!\downarrow) \, \frac{K_f(\uparrow\!\downarrow\to \uparrow\!\uparrow)}{K_f(\uparrow\!\downarrow\to \downarrow\!\downarrow) + K_f(\uparrow\!\downarrow\to \uparrow\!\uparrow)} \\
			&+ K_s(\downarrow\!\downarrow \to \downarrow\!\uparrow) \, \frac{K_f(\downarrow\!\uparrow \to \uparrow\!\uparrow)}{K_f(\downarrow\!\uparrow \to \uparrow\!\uparrow) + K_f(\downarrow\!\uparrow \to \downarrow\!\downarrow)} \comma
	}
	\end{equation}
	and
	\begin{equation}
	\eqalign{
		K_c (A\to I) = K_s&(\uparrow\!\uparrow \to \downarrow\!\uparrow) \, \frac{K_f(\downarrow\!\uparrow \to \downarrow\!\downarrow)}{K_f(\downarrow\!\uparrow \to \uparrow\!\uparrow) + K_f(\downarrow\!\uparrow \to \downarrow\!\downarrow)} \\
			&+ K_s(\uparrow\!\uparrow \to \uparrow\!\downarrow) \, \frac{K_f(\uparrow\!\downarrow \to \downarrow\!\downarrow)}{K_f(\uparrow\!\downarrow \to \downarrow\!\downarrow) + K_f(\uparrow\!\downarrow \to \uparrow\!\uparrow)} \fs
	}
	\end{equation}
	By substituting Eqs.\,(\ref{eq:glauber-fast}) and (\ref{eq:glauber-slow}) into these expressions, we find
	\begin{equation}
		K_c(A\to I,\,\{\ell_i\}) = \omega_f\,\Bigg\{
			\frac{e^{-h_1-\lambda_1}}{1 + e^{h_1 +h_2 - \lambda_1 + \lambda_2}} +
			\frac{e^{-h_2-\lambda_2}}{1 + e^{h_1 +h_2 + \lambda_1 - \lambda_2}}
		\Bigg\}			
	\end{equation}
	and
	\begin{equation}
		K_c(I\to A,\,\{\ell_i\}) = \omega_f\,\Bigg\{
			\frac{e^{h_1-\lambda_1}}{1 + e^{-(h_1 +h_2) - \lambda_1 + \lambda_2}} +
			\frac{e^{h_2-\lambda_2}}{1 + e^{-(h_1 +h_2) + \lambda_1 - \lambda_2}}
		\Bigg\} \fs
	\end{equation}
	Even in this simple case with 2 protomers, the first time-scale separation yields effective rates which depend on the full binding state in a non trivial way, and not only on the sum $l = \sum \ell_i$.
%	We have shown that, strictly speaking, the reduced model is not equivalent to the MWC model, in which the states are specified only by the activity state $A$ or $I$, and the total number $l$ of bound protomers.
	
	However,  being detailed balance preserved by the coarse-graining procedure, we have
	\begin{equation}
			\frac{K_c(A\to I,\,\{\ell_i\})}{K_c(I\to A,\,\{\ell_i\})} =
			\frac{P_{eq}(I|\{\ell_i\})}{P_{eq}(A|\{\ell_i\})} = e^{-2 (h_1 + h_2)} \comma
	\end{equation}
	consistently with the general formula~\footnote{The spin variable $\sigma$ in this expression corresponds to the  spin-activity variable of the whole coherent system.}
	\begin{equation}
		P_{eq}(\sigma|\{\ell_i\}) = \frac{1}{Z(\{\ell_i\})}\,e^{\sigma\,\sum_i h_i} \comma 
	\end{equation}
	which gives the Boltzmann weights according to the Hamiltonian \eqref{eq:CS-ham} restricted to the coherent configuration (in which case the coupling term is a constant contribution cancelled by the normalization $Z$).
	From Eq.\,\eqref{eq:magfield}, it is obvious that such Boltzmann weights only depend on the global occupancy $l = \sum \ell_i$.

	\subsection*{$N$-protomer, completely unbound case}	

	In the general case with $N$ protomers, all unbound~\footnote{In general, with all the protomers with the same occupancy.}, we are able to map the coarsening process into a simple birth--and--death process which, in the limit $\gamma \to 1$, has site-independent rates.
	First of all, as we remarked in the main text, because of the time-scale separation, the fast coarsening process does not involve any state but the coherent (which are the long-living states) and the ones with only two domain walls.
	Then, the coarsening is simply a motion of the domain walls, namely an expansion or contraction of the domain which has been nucleated, until one of the coherent states is reached.
	The expansion/contraction of this domain can happen by a flip of a spin at its right or left border:
	if the protomer occupancy $\ell$ is the same for all sites, it is not important at which side of the domain the expansion/contraction occurs.
	Therefore, we can label the states visited by the coarsening process just by the number of protomers in the, \emph{e.g.}, active state (spin up), denoted by $n$:
	far from the absorbing states $n=0$ and $n=N$, the rate of increasing or decreasing the number of active protomers is twice the rate of moving a domain wall;
	the absorption rates from the state $n=1$ and $n=N-1$, are the rates of absorbing the two domain walls.
	The process is schematically represented here,
	\begin{center}
		\includegraphics[width=\textwidth]{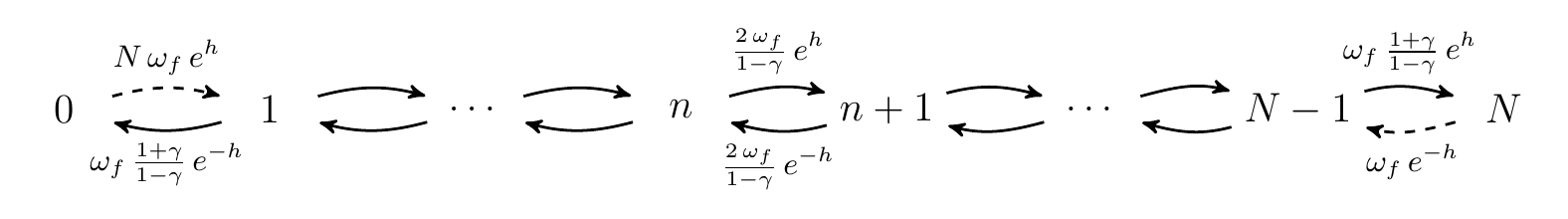}
	\end{center}
	where the (reduced) magnetic field $h$ is the value of $h_i$ given by Eq.\,(\ref{eq:magfield-2}) specified to $\ell_i = 0$, so
	\[
		h = \frac{\beta}{2}(\varepsilon_I - \varepsilon_A) \smc
	\]
	such (negative) value is responsible for the downward alignment of the spin-activity variables, namely favouring the inactive state.
	The solid arrows represent the fast rates, while the dashed ones are the slow rates of nucleation of one domain.

	In the strong-coupling limit, $\gamma \to 1$, and all the fast rates are asymptotically equal;
	the coarsening dynamics is then formally described by a birth-and-death process with site-independent rates, defined on the integer numbers between $0$ and $N$, \emph{i.e.} as asymmetric random walk with absorbing boundary conditions.
	The probability of being absorbed in the state $N$, starting from the state 1, is easily calculated to be
	\begin{equation}
		P_{abs}(N\,|\,1) = e^{h\,(N-1)}\,\frac{\sinh h}{\sinh Nh} \fs
	\end{equation}
	Once multiplied by the slow exit rate from 0 to 1, this gives the effective rate of switching from the inactive to the active state, in absence of ligands:
	\begin{equation}\label{eq:on_empty}
		K_c(I \to A,\,\{\ell_i = 0\}) = N\,\omega_f\,e^{h\,N}\,\frac{\sinh h}{\sinh N h} = N\,\omega_f\,L^{-1/2}\,\frac{\sinh h}{\sinh N h}  \comma
	\end{equation}
	where $L$ is the allosteric constant of the $N$-protomers MWC molecule, defined in the main text as $L = (k_i/k_a)^N = \exp N\beta(\varepsilon_A - \varepsilon_I)$.
	Similarly, for the opposite switch, one has
	\begin{equation}\label{eq:off_empty}
		K_c(A \to I,\,\{\ell_i = 0\}) = N\,\omega_f\,e^{-h\,N}\,\frac{\sinh h}{\sinh N h} = N\,\omega_f\,L^{1/2}\,\frac{\sinh h}{\sinh N h}  \fs
	\end{equation}
	These rates are exact up to a correction of order $1-\gamma$, which is exponentially small in the coupling $\beta J$ (see Sec.\,\ref{sec:CS}).
	We notice that the ratio between these switching rates is the allosteric constant $L$, as expected from the detailed balance condition and from the Hamiltonian \eqref{eq:CS-ham}.
	The one in Eq.\,(\ref{eq:on_empty}), or, alternatively, Eq.\,(\ref{eq:off_empty}), is the overall frequency scale of the switching dynamics in the MWC model.
	All the other switching rates are found from these ones by applying the detailed balance condition.

	\pagebreak
	\section*{References}
%	\nocite{*}

\end{document}